\documentclass[12pt]{amsart}
\usepackage[cp1251]{inputenc}
\usepackage[english]{babel}
\usepackage{amsmath}
\usepackage{amssymb}
\usepackage{amsfonts}
\usepackage{epsfig}
\usepackage{srcltx}
\usepackage{subfigure}
\usepackage{cite}
\usepackage{float}
\usepackage[ a4paper,  mag=1000, includefoot,  left=2cm, right=2cm, top=2cm, bottom=2cm, headsep=1cm, footskip=1cm ]{geometry}
\usepackage[unicode,colorlinks]{hyperref}
\hypersetup{colorlinks=true, citecolor=blue, linkcolor=blue}

\newtheorem{Th}{Theorem}

\newtheorem{Rem}{Remark}
\begin{document}
\thispagestyle{empty}

\title[Stabilization of unstable autoresonant modes]
{Stabilization of unstable autoresonant modes}

\author{Oskar Sultanov}

\address{Oskar Sultanov,
\newline\hphantom{iii} Institute of Mathematics, Ufa Scientific Center, Russian Academy of Sciences,
\newline\hphantom{iii}  112, Chernyshevsky str., Ufa, Russia, 450008}
\email{oasultanov@gmail.com}


\maketitle {\small
\begin{quote}
\noindent{\bf Abstract.}
A mathematical model describing the initial stage of the capture of oscillatory systems into autoresonance under the action of slowly varying pumping is considered. Solutions with an infinitely growing amplitude are associated with the autoresonance phenomenon. Stability of such solutions is of great importance because only stable solutions correspond to physically observable motions. We study the stabilizing problem and we show that the adiabatically varying parametric perturbation with decreasing amplitude in time can stabilize the unstable autoresonant modes.
\medskip


\end{quote}
}

\section*{Introduction}

Autoresonance is a phenomenon that occurs when a forced nonlinear oscillator stays phase-locked with a slowly changing periodic pumping. The essence of this phenomenon is that the oscillator automatically adjusts to the pumping and remains in the resonance for a long time interval. This can leads to a significant growth of the energy of the oscillator. It is considered that such a phenomenon has been found in the problems associated with the acceleration of relativistic particles~\cite{Veks45,McM45}. Later it was found that autoresonance plays an important role in many problems of nonlinear physics~\cite{FajFried01}. The mathematical theory of the autoresonance is based on a study of differential equations with slowly varying perturbations whose solutions describe the resonant behavior of nonlinear oscillators~\cite{Neish75}. The initial stage of the capture in autoresonance for different nonlinear systems is described by a number of mathematical models~\cite{LK08} whose solutions with growing energy or amplitude are associated with the autoresonance. The existence and stability of such solutions are of great importance because only stable solutions correspond to the observable states of physical systems.

The mathematical model of the autoresonance in systems with the external pumping possess two autoresonant modes with a different phase behaviour. From~\cite{LKOS13} it follows that one of them is stable while the other is unstable. In this paper we study the problem of stabilization and we prove that the small adiabatically varying parametric pumping can stabilize the unstable autoresonant solutions. In particular, we determine the range of the parameters that guarantee the simultaneous stability of both autoresonant modes. These results extend the possibility of using the autoresonance phenomenon for sustainable excitation and control of nonlinear systems.

The paper is organized as follows. In section 1, we give the mathematical formulation of the problem. Section 2 deals with asymptotics for the autoresonance modes. Section 3 provides linear stability analysis of autoresonant solutions. Nonlinear stability analysis is contained in section 4. The paper concludes with a brief discussion of the results obtained.

\section{Problem statement}
We study the system of two differential equations
\begin{gather}
\label{PRst0}
        \frac{d\rho}{d\tau}= f \sin\psi, \quad \rho\Big[\frac{d\psi}{d\tau}-\rho^2+ \lambda \tau\Big]= f \cos\psi, \quad \tau>0
\end{gather}
with the parameters $\lambda,f\neq 0$. This system appears after the averaging of a wide class of equations describing the behaviour of  nonlinear oscillatory systems in the presence of a slowly varying external force and describes the dynamics of the principal terms of the asymptotic behavior of the resonant solutions. Since the system is invariant under replacements of $f$ and $\psi$ with $-f$ and $\psi+\pi$ respectively, we can assume that $f>0$. The unknown functions $\rho(\tau)$ and $\psi(\tau)$ play the role of the amplitude and the phase shift of harmonic oscillations. The solutions with unboundedly growing amplitude $\rho(\tau)\to\infty $ and limited phase shift $\psi(\tau)=\mathcal O(1)$ as $\tau \to\infty$ are associated with the capture of oscillatory system into the autoresonance. Such solutions can exist only if $\lambda>0$. We assume that this condition holds.

Asymptotic analysis and numerical simulations show that system \eqref{PRst0} has two types of captured solutions with growing amplitude and different behavior of the phase variable~\cite{LF08,LK08,LKOS13}. The solutions with $\rho(\tau)\approx \sqrt{\lambda \tau}$ and $\psi(\tau)\approx \pi$ as $\tau\to\infty$ are stable, while the solutions with $\psi(\tau)\approx 0$ are unstable.

In order to stabilize of unstable autoresonant solutions we consider the perturbed system in the form
\begin{gather}
\label{PRst}
        \frac{d\rho}{d\tau}+\nu(\tau) \rho \sin 2\psi= f \sin\psi, \quad \rho\Big[\frac{d\psi}{d\tau}+ \nu(\tau) \cos 2\psi-\rho^2+ \lambda \tau\Big]= f \cos\psi, \quad \tau>0
\end{gather}
where $\nu(\tau)= m/(1+\tau)^{1/2}$, $m={\hbox{\rm const}}\neq 0$. The perturbation is associated with the slowly varying parametric pumping. Note that system \eqref{PRst} with $\nu(\tau)\equiv 1$ and $f=0$ was considered in~\cite{LKOS13,OS16}, where it was proved the existence and stability of autoresonant solutions with $\psi(\tau)\to 0$ and $\psi(\tau)\to \pi$. It was shown in~\cite{OK16} that the external excitation with decreasing amplitude ($f=\mathcal O(\tau^{-1/2})$, $\nu\equiv 0$) can lead to autoresonant capture of nonlinear systems. The ability of using parametric pumping with decreasing amplitude ($\nu(\tau)\to 0$ as $\tau\to\infty$) to excite and to stabilize of autoresonant modes has not been considered.

In this paper, we construct asymptotics for isolated autoresonance solutions to system \eqref{PRst} as $\tau\to\infty$ and we provide a careful stability analysis of these solutions with respect to initial data perturbations. Moreover, we show that these solutions attract the set of the captured solutions. Finally, the stabilizing effect of unstable autoresonant solutions to system \eqref{PRst0} by decreasing parametric perturbations is discussed.

\begin{figure}
\begin{minipage}[h]{0.47\linewidth}
\center{\includegraphics[width=1\linewidth]{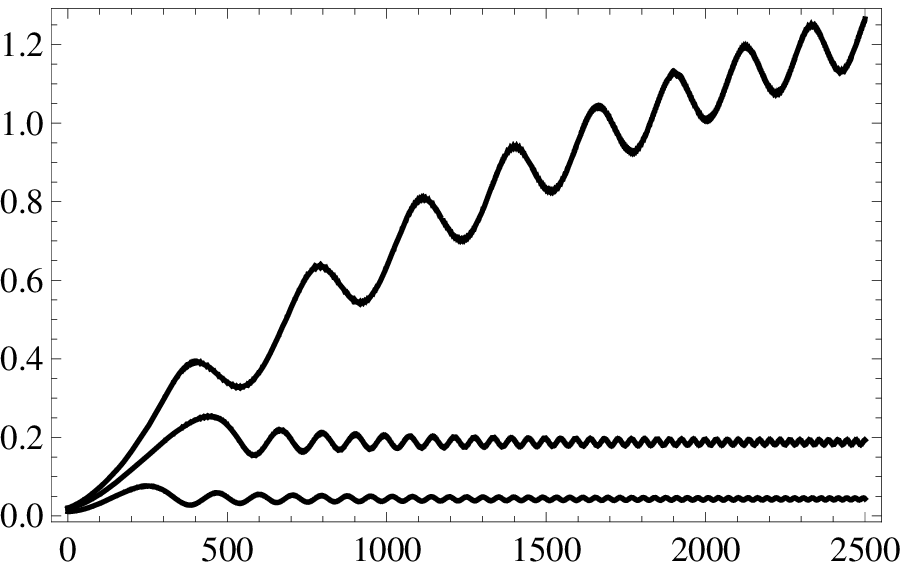}} a \\
\end{minipage}
\hfill
\begin{minipage}[h]{0.47\linewidth}
\center{\includegraphics[width=1\linewidth]{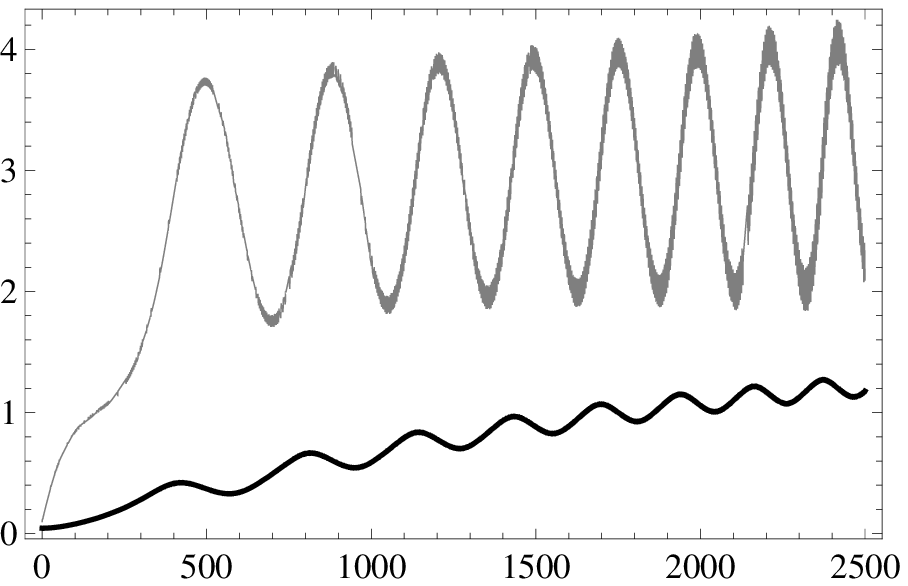}}  b \\
\end{minipage}
\vfill
\begin{minipage}[h]{0.47\linewidth}
\center{\includegraphics[width=1\linewidth]{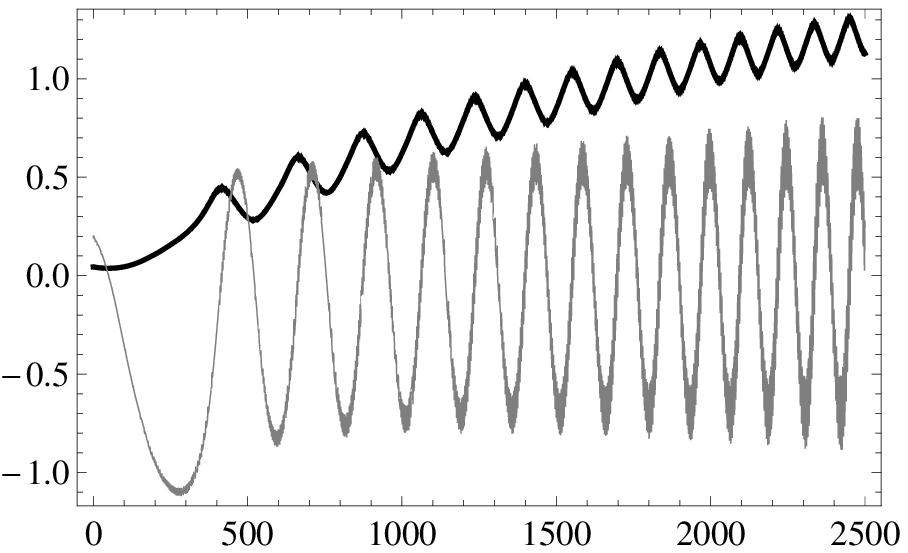}} c \\
\end{minipage}
\hfill
\begin{minipage}[h]{0.47\linewidth}
\center{\includegraphics[width=1\linewidth]{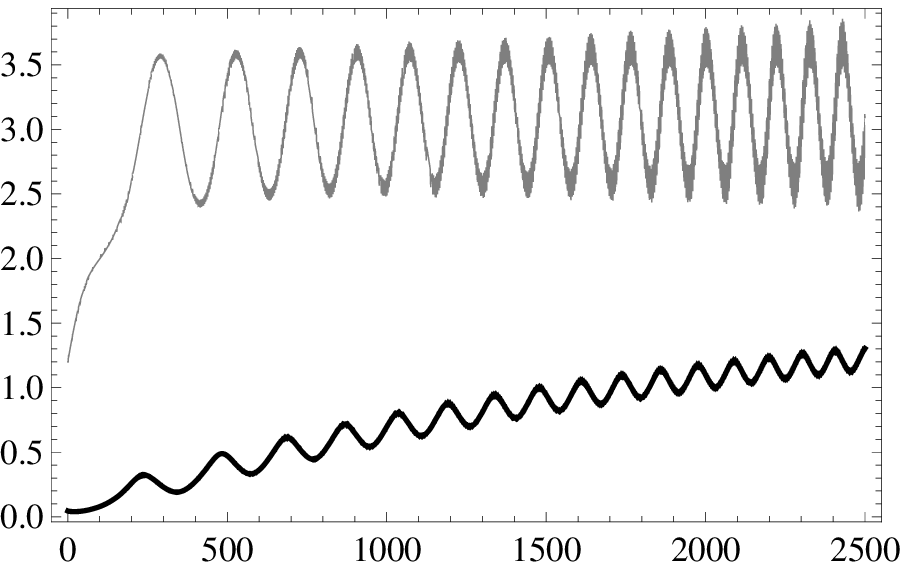}} d \\
\end{minipage}
\caption{The evolution of the energy $E(t)$ (black curve) and the phase mismatch $\Delta(t)$ (gray curve) for solutions of \eqref{EPst} with different initial data, $\varepsilon=10^{-3}$, $\alpha=10^{-4}/2$, $\gamma=1/6$, (a,b):  $f_0=4$, $h_0=0$; (c,d):  $f_0=1$, $h_0=5$.}
\label{Pic1}
\end{figure}

System \eqref{PRst0} is of universal character in the mathematical description of autoresonance in nonlinear systems. It describes long-term evolution of nonlinear oscillations under the action of small slowly varying force. As but one example let us consider the equation:
\begin{gather}
\label{EPst}
        \frac{d^2u}{dt^2}+\Big(1+\varepsilon^{2/3} h(t) \cos 2\phi(t)\Big)U'(u)=\varepsilon f_0 \cos\phi(t),
\end{gather}
where $U(u)=u^2/2-\gamma u^4/4+\mathcal O(u^6)$ as $u\to 0$, $\phi(t)=t-\alpha t^2$, $h(t)=h_0 /(1+\varepsilon^{2/3}t)^{1/2}$, $0<\varepsilon,|\alpha|\ll 1$, $f_0,h_0,\gamma={\hbox{\rm const}}\neq 0$, $\gamma>0$. It is easy to see that the equation with $\varepsilon=0$ has locally stable fixed point $u=0$, $u'=0$. The solutions starting from a small neighbourhood of the equilibrium $|u(0)|+|u'(0)|=\mathcal O(\varepsilon^{1/3})$ whose the energy $E(t)=U\big(u(t)\big)+\big(u'(t)\big)^2/2$ increases to the order of unity and the phase $\Phi(t)=-\arctan \big(u'(t)/u(t)\big)$ is synchronised with the pumping such that $\Delta(t)=\phi(t)-\Phi(t)=\mathcal O(1)$ are associated with the autoresonance phenomenon (see. Fig.~\ref{Pic1}, a). Note that equation \eqref{EPst} with $h(t)\equiv 0$, $f_0>0$, and $\alpha>0$ has anti-phase autoresonant solutions with $\Delta(t)\sim \pi$, while the in-phase autoresonance with $\Delta(t)\sim 0$ does not occur (see Fig.~\ref{Pic1}, b). However, the coexistence of two autoresonant modes with $\Delta(t)\sim 0$ and $\Delta(t)\sim \pi$ for different initial data is possible if $h_0\neq 0$, $f_0>0$, $\alpha>0$ and some additional conditions hold (see Fig.~\ref{Pic1}, c, d). Thus, the decreasing parametric pumping can stabilize the unstable autoresonant modes. For asymptotic description of autoresonant solutions to equation \eqref{EPst}, it is convenient to use the method of two scales~\cite{BogMitr61}. We introduce a slow time $\tau=\varepsilon^{2/3}t$ and a fast variable $\phi=\phi(t)$, then the asymptotic substitution
$u(t)= \varepsilon^{1/3}\rho(\tau) \cos \big(\psi(\tau)-\phi(t)\big)\sqrt{{8}/{3\gamma}}  +\mathcal O(\varepsilon)$
in equation \eqref{EPst} and the averaging procedure over the fast variable $\phi(t)$ in the leading-order term in $\varepsilon$ lead to system \eqref{PRst} for the slowly varying functions $\rho(\tau)$ and $\psi(\tau)$, where $\lambda=2\alpha \varepsilon^{-4/3}$, $ m=h_0/4$, $f=f_0\sqrt{3\gamma / 32}$. In the case $h(t)\equiv 0$, this procedure leads to system \eqref{PRst0}.

\section{Asymptotics of autoresonant solutions}
The asymptotic solution of \eqref{PRst} with unboundedly growing amplitude $\rho(\tau)$ can be constructed in the form
\begin{gather}
\label{ASstep}
    \rho_\ast(\tau)=\sqrt {\lambda \tau} + \rho_0 +\sum_{k=1}^\infty \rho_k \tau^{-k/2}, \quad  \psi_\ast(\tau)=\psi_0 +\sum_{k=1}^\infty \psi_k \tau^{-k/2}.
\end{gather}
Substituting these series in \eqref{PRst} and grouping the expressions of the same powers of $\tau$ give the recurrence relations for determining the constant coefficients $\rho_k$, $\psi_k$.
Following this approach, we find few asymptotic solutions $\rho_\ast^i(\tau)$, $\psi_\ast^i(\tau)$, distinguished by the choice of a root to the equation:
$$
    ( m_{\ast}-  m\cos\psi_0)\sin\psi_0 =0, \quad  m_{\ast}:=\frac{f}{\sqrt{4\lambda}}>0.$$
Note that in case of $| m|> m_{\ast}$, the trigonometric equation for $\psi_0$ has four different roots:
\begin{gather}
\label{psi0}
\psi_0^1=0,\quad \psi_0^2=\pi, \quad \psi_0^3=\arccos\frac{ m_{\ast}}{ m}, \quad \psi_0^4=-\arccos\frac{ m_{\ast}}{ m}.
\end{gather}
In case of $| m|< m_{\ast}$, the equation has only two roots: $\psi_0^1=0$ and $\psi_0^2=\pi.$
In both cases we have
$$
\rho_0=\rho_1=0,\ \ \rho_2=\frac{ m\cos2\psi_0-2 m_{\ast}\cos\psi_0}{\sqrt{4\lambda}}, \ \ \psi_1=\frac{1}{4 m_{\ast}\cos\psi_0 - 4  m\cos 2\psi_0},
$$
etc. Note that in case of $ m=\pm  m_{\ast}$, the trigonometric equation for $\psi_0$ also has two roots, while the other coefficients $\rho_k$, $\psi_k$ are no longer defined. In this case, there is no asymptotic solution in the form \eqref{ASstep}.

\begin{Th}
If $|m|>f/\sqrt{4\lambda}$, then system \eqref{PRst} has 4 solutions with asymptotic expansion in the form of a series \eqref{ASstep} with $\psi_0\in\{0,\pi,\arccos (m_{\ast}/m),-\arccos (m_{\ast}/m)\}$, where $m_{\ast}=f/\sqrt{4\lambda}$.
If $|m|<f/\sqrt{4\lambda}$, then system \eqref{PRst} has 2 solutions with asymptotic expansion in the form of a series \eqref{ASstep} with $\psi_0\in\{0,\pi\}$.
\end{Th}

The existence of solutions $\rho_\ast^i(\tau)$, $\psi_\ast^i(\tau)$ with the asymptotics \eqref{ASstep} as $\tau\geq \tau_\ast>0$ follows from~\cite{Kuzn89,KozFur13}.
It follows from the comparison theorems (see~\cite{Halil,LK14}) that these solutions can be extended to the whole real line.
We discuss the stability of solutions $\rho^i_\ast(\tau)$, $\psi^i_\ast(\tau)$ as $\tau \to \infty$.
Note that the structure of the capture region for system \eqref{PRst} remains unknown, its description is beyond the scope of this paper.
This region was analyzed in~\cite{OKNT14} for a similar model.

\begin{figure}
\vspace{-2ex} \centering \subfigure[]{
\includegraphics[width=0.4\linewidth]{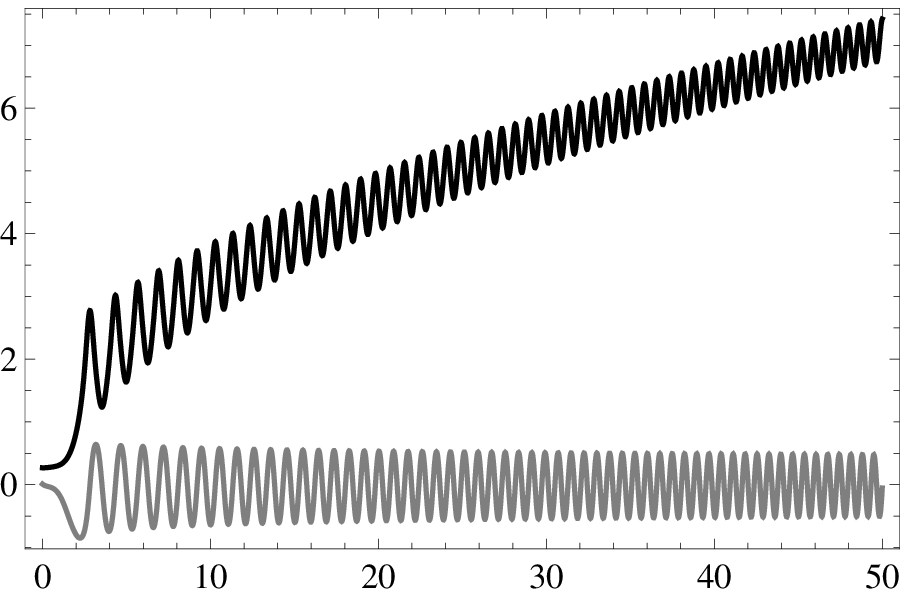} \label{fig:a}
}
\hspace{4ex}
\subfigure[]{
\includegraphics[width=0.4\linewidth]{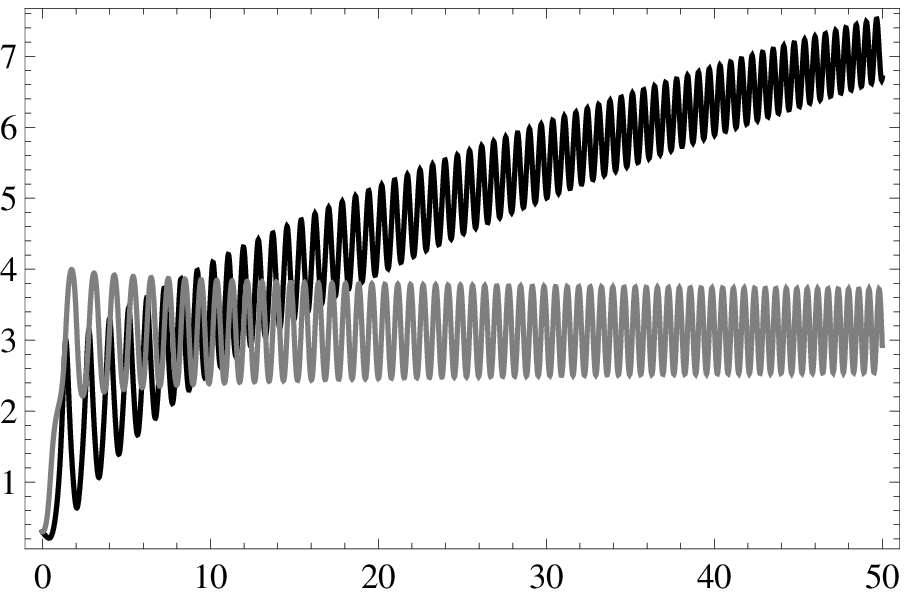} \label{fig:b}
}
\hspace{4ex}
\subfigure[]{ \includegraphics[width=0.4\linewidth]{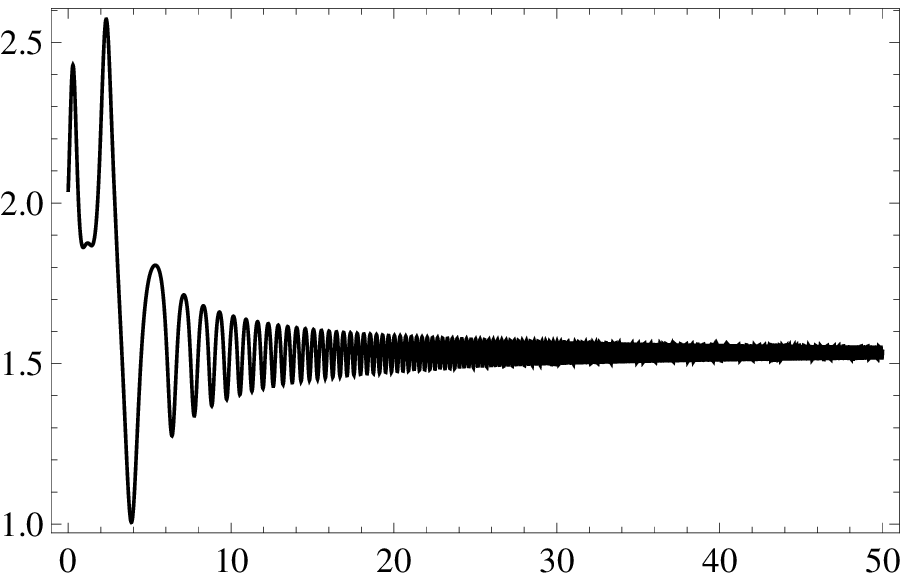} \label{fig:c} }
\caption{The evolution of the energy $\rho(\tau)$ (black curve) and phase shift $\psi(\tau)$ (gray curve) for solutions of \eqref{PRst} with $\lambda=1$, $f=1$, $m=4$ and different initial data: \subref{fig:a}  $\rho(0)=0.27$, $\psi(0)=0.01$; \subref{fig:b} $\rho(0)=0.32$, $\psi(0)=0.31$; \subref{fig:c} $\rho(0)=2.04$, $\psi(0)=1.78$.} \label{Pic2}
\end{figure}

We note that system \eqref{PRst} also has many autoresonant solutions with more complicated asymptotic behaviour at infinity (see Fig.~\ref{Pic2}, a, b) and non-resonant solutions with bounded energy (see Fig.~\ref{Pic2}, c).  In this paper, we focus only on the autoresonant solutions with power asymptotics.

\section{Linear stability analysis}
Linear stability analysis allows to determine the character of instability of the autoresonant solutions. In system \eqref{PRst}, we make the change of variables
$$\rho(\tau)=\rho^i_\ast(\tau) + r(\eta) \tau^{-1/4}, \quad \psi(\tau)=\psi_\ast^i(\tau)+p(\eta), \quad \eta=\frac{4}{5} \tau^{5/4}$$
and we study the stability of the equilibrium $(0,0)$ in the system
\begin{gather}
\frac{dr}{d\eta}=-\partial_p H^i(r,p,\eta), \quad \frac{dp}{d\eta}=\partial_r H^i(r,p,\eta)+ G^i(r,p,\eta),
    \label{Sys0}
\end{gather}
where
\begin{eqnarray*}
  H^i &=& \frac{r^2 \rho^i_\ast}{\sigma^2 \eta^{2/5}}+\frac{r^3}{3\sigma^3 \eta^{3/5}}+ f[\cos(p+\psi^i_\ast)-\cos\psi^i_\ast+p\sin\psi^i_\ast] - \frac{rp}{5\eta} -  \\
            & & -\frac{\nu\rho^i_\ast}{2}\Big[\cos(2p+2\psi^i_\ast)-\cos2\psi^i_\ast+2p \sin2\psi^i_\ast\Big] - \frac{\nu r}{2\sigma\eta^{1/5}}\Big[\cos(2p+2\psi^i_\ast)-\cos2\psi^i_\ast\Big],  \\
   G^i &=& -\frac{\nu}{2\sigma\eta^{1/5}}\Big[\cos(2p+2\psi^i_\ast)-\cos2\psi^i_\ast\Big]+\frac{f}{\sigma \eta^{1/5}}\Big[\frac{\cos(p+\psi^i_\ast)}{\rho_\ast^i+r/\sigma \eta^{1/5}}-\frac{\cos\psi^i_\ast}{\rho^i_\ast}\Big]+\frac{p}{5\eta},
\end{eqnarray*}
$\sigma=(5/4)^{1/5}$. Consider the linearized system in the vicinity of the fixed point $(0,0)$:
\begin{gather}
    \frac{d}{d\eta}\begin{pmatrix} r \\ p \end{pmatrix} = {\bf A}^i(\eta) \begin{pmatrix} r \\ p \end{pmatrix}, \quad {\bf A}^i(\eta)=\begin{pmatrix} \displaystyle \frac{1}{5\eta}-\frac{\nu \sin2\psi_\ast^i}{\sigma\eta^{1/5}} & \displaystyle f\cos\psi_\ast^i -   2 \nu \rho_\ast^i  \cos2\psi_\ast^i \\ \displaystyle \frac{2\rho_\ast^i}{\sigma^2\eta^{2/5}}-\frac{f\cos\psi_\ast^i}{(\rho_\ast^i)^2 \sigma^2\eta^{2/5}}  & \displaystyle \frac{2\nu\sin2\psi_\ast^i}{\sigma\eta^{1/5}}-\frac{f\sin\psi_\ast^i}{\sigma \eta^{1/5}\rho_\ast^i} \end{pmatrix}.
    \label{Sys}
\end{gather}
Denote by $\mu_\pm^i(\eta)=\delta^i(\eta) \pm\sqrt{D^i(\eta) }$, $i\in \{1,2,3,4\}$ the roots of the characteristic equation $|{\bf A}^i-\mu {\bf I}|=0$. Then
\begin{eqnarray*}
  \delta^i(\eta) &=& \frac{( m\cos\psi_0^i- m_{\ast})\sin\psi_0^i}{\sigma^3 \eta^{3/5}}+\mathcal O(\eta^{-1}), \\
  D^i(\eta) & = & 4  ( m_{\ast}\cos\psi_0^i -  m\cos2\psi_0^i)\lambda+\mathcal O(\eta^{-2/5}) \quad {\rm as} \quad  \eta\to\infty.
\end{eqnarray*}
Using \eqref{psi0}, we get the asymptotics of eigenvalues: $\mu_\pm^i(\eta)=\pm \sqrt{D_0^i}+\mathcal O(\eta^{-2/5})$, where
\begin{gather*}
   D_0^1=4 ( m_{\ast}- m)\lambda,\quad D_0^2=-4 ( m_{\ast}+ m)\lambda,\quad
  D_0^3=D_0^4=\frac{4( m^2- m_{\ast}^2)\lambda}{ m}.
\end{gather*}
If $D_0^i>0$, then the leading asymptotic terms of eigenvalues are real of different signs. This implies that the fixed point $(0,0)$ of \eqref{Sys} and the corresponding autoresonance solution $\rho_\ast^i(\tau)$, $\psi_\ast^i(\tau)$ to system \eqref{PRst} are unstable at infinity. In the opposite case, when $D_0^i<0$, the leading asymptotic terms of eigenvalues are pure imaginary and $\Re[\mu_\pm^i(\eta)]\to 0$ as $\eta\to\infty$. In this case, the linear stability analysis fails (see~\cite{LKOS13} and \cite[p.~84]{Demid}), and the stability property depends on nonlinear terms of equations. Thus we have
\begin{Th}
The solution $\rho_\ast^1(\tau)$, $\psi_\ast^1(\tau)$ is unstable if $m>f/\sqrt{4\lambda}$,.
The solution $\rho_\ast^2(\tau)$, $\psi_\ast^2(\tau)$ is unstable whenever $m>-f/\sqrt{4\lambda}$ and $m\neq f/\sqrt{4\lambda}$.
The solutions $\rho_\ast^3(\tau)$, $\psi_\ast^3(\tau)$ and $\rho_\ast^4(\tau)$, $\psi_\ast^4(\tau)$ are unstable if $m<-f/\sqrt{4\lambda}$.
\label{Th01}
\end{Th}

The table~\ref{Tab1} shows the intermediate results on the stability of autoresonant solutions $\rho_\ast^i(\tau)$, $\psi_\ast^i(\tau)$.
Let us remark that the autoresonance mode with $\psi\sim 0$ is unstable in the system without parametric perturbation ($m=0$). However, its stabilization is possible if $m>m_{\ast}$. On the other hand, if $m<-m_{\ast}$, then both autoresonant solutions with $\psi\sim 0$ and $\psi\sim \pi$ are unstable, while the other pair of captured solutions with $\psi\sim \psi_0^3$ and $\psi\sim \psi_0^4$ may be stable. The stability of these solutions is discussed in the next section.

\begin{table}[H]
\caption{Stability of solutions $\rho_\ast^i(\tau)$, $\psi_\ast^i(\tau)$}
\label{Tab1}
\begin{tabular}{c|c|c|c}
\hline         & $ m<- m_{\ast}$ & $- m_{\ast}< m< m_{\ast}$ & $ m> m_{\ast}$  \\
\hline
$i=1$ & unstable  & unstable & ? \\
\hline
$i=2$ & unstable  & ?  & ? \\
\hline
$i=3$ & ?  & -  & unstable \\
\hline
$i=4$ & ?  &   - & unstable \\
\hline
\end{tabular}
\end{table}

\section{Nonlinear stability analysis}

In this section, we study the stability of solutions $\rho_\ast^i(\tau)$, $\psi_\ast^i(\tau)$, $i\in \{1,2,3,4\}$ by Lyapunov's second method. In the analysis, we use the leading terms of the asymptotics with the remainder estimate
\begin{gather}
  \rho_\ast^i(\tau)=\sqrt {\lambda \tau} +\rho_2^i \tau^{-1}+\mathcal O(\tau^{-3/2}), \quad  \psi_\ast^i(\tau)=\psi_0^i+\psi_1^i \tau^{-1/2}+\mathcal O(\tau^{-1}), \quad as\quad \tau\to\infty.
  \label{Asest}
  \end{gather}
We have
\begin{Th}
If $m>f/\sqrt{4\lambda}$, then the solutions $\rho_\ast^1(\tau)$, $\psi_\ast^1(\tau)$ and $\rho_\ast^2(\tau)$, $\psi_\ast^2(\tau)$ are asymptotically stable.  Moreover, for all $i\in\{1,2\}$ and $0<s<1/8$ there exist $\delta_0>0$ and $\tau_0>0$ such that for all $(R_0, \Psi_0){\rm :}$ $ (R_0-\rho^i_\ast(\tau_0))^2+(\Psi_0-\psi^i_\ast(\tau_0))^2\leq \delta_0^2$ the solution $\rho(\tau)$, $\psi(\tau)$ to system \eqref{PRst} with initial data $\rho(\tau_0)=R_0$, $\psi(\tau_0)=\Psi_0$ has the asymptotics
\begin{gather}
\rho(\tau)=\sqrt{\lambda \tau}+\mathcal O(\tau^{-1/4-s}), \quad \psi(\tau)=\psi_0^i+\mathcal O(\tau^{-s}) \quad {\rm as }\quad \tau>\tau_0,
\label{AsEst}
\end{gather}
where $\psi_0^1=0$ and $\psi_0^2=\pi$.
\label{Th1}
\end{Th}
\proof
In system \eqref{PRst} we make the change of variables $\rho(\tau)=\rho^i_\ast(\tau) + r(\eta)\tau^{-1/4}$, $\psi(\tau)=\psi_\ast^i(\tau)+p(\eta)$, $\eta=4\tau^{5/4}/5$, $i\in\{1,2\}$ and for new functions $r(\eta)$, $p(\eta)$ we study the stability of the equilibrium $(0,0)$ of system \eqref{Sys0} as $\eta\to\infty$.

To construct a Lyapunov function for system \eqref{Sys0} we use the asymptotics of the right-hand sides at infinity (as $\eta\to\infty$) and in a neighborhood of the equilibrium (as $d:=\sqrt{r^2+p^2}\to 0$). Notice that all asymptotic estimates written out bellow in the form $\mathcal O(\eta^{-\kappa})$ and $\mathcal O(d^\kappa)$, ($\kappa={\hbox{\rm const}}>0$) are uniform with respect to $(r,p,\eta)$ in the domain $\mathcal D (d_\ast,\eta_\ast)=\{ (r,p,\eta)\in\mathbb R^3: d<d_\ast, \eta>\eta_\ast \}$, where $d_\ast,\eta_\ast={\hbox{\rm const}}>0$.
By taking in account \eqref{Asest} one can readily write out the asymptotics of the derivatives of the Hamiltonian:
\begin{eqnarray*}
 \partial_p H^i  &=& \frac{f  m}{2 m_{\ast}}  \sin 2p -f \cos\psi_0^i \sin p +  \\
    & & + \frac{f\psi_1^i}{\sigma^2 \eta^{2/5}} \Big[\cos\psi_0^i (1-\cos p)+\frac{ m}{ m_{\ast}}  (\cos 2p - 1)\Big]  + \mathcal O(\eta^{-3/5})\mathcal O(d), \\
 \partial_r H^i &=& 2\sqrt\lambda r+\mathcal O(\eta^{-3/5}) \mathcal O(d), \\
 \partial_\eta H^i &=& \mathcal O(\eta^{-7/5})\mathcal O(d^2).
\end{eqnarray*}
The non-Hamiltonian part $G^i(r,p,\eta)$ tends to zero as $\eta\to\infty$:
\begin{eqnarray*}
              G^i &=&  \frac{g^i(p)}{\sigma^3 \eta^{3/5}} - \frac{1}{5 \eta} [p+ \mathcal O(d^2)]  + \mathcal O(\eta^{-6/5})\mathcal O(d),\\
              g^i(p) &:=& \frac{ m}{2} (1-\cos 2p)-2 m_{\ast}\cos\psi_0^i (1-\cos p).
\end{eqnarray*}
It can easily be checked that if $ m>m_{\ast}=f/\sqrt{4\lambda}$, the Hamiltonian has a positive quadratic form as the leading term of the asymptotic expansion:
\begin{eqnarray*}
  H^i   &=& \sqrt\lambda r^2  + \frac{f m}{4 m_{\ast}}(1-\cos 2p)- f \cos\psi_0^i (1-\cos p)+\\
                & & +\frac{f \psi_1^i}{\sigma^2 \eta^{2/5}}\Big[ \cos\psi_0^i(p- \sin p)  + \frac{ m}{2  m_{\ast}} (\sin 2p-2p)  \Big]+ \mathcal O(\eta^{-3/5})\mathcal O(d^2) = \\
        & = & \sqrt{\lambda} r^2+\frac{ ( m- m_{\ast}\cos\psi_0^i)  f p^2}{2 m_{\ast}}+\mathcal O(d^3)+\mathcal O(\eta^{-2/5})\mathcal O(d^2)
\end{eqnarray*}
as $\eta\to\infty$ and $d\to 0$.
A Lyapunov function candidate for system \eqref{Sys0} is constructed of the form
$$L^i(r,p,\eta) = H^i(r,p,\eta) + V^i_1(r,p,\eta)+V^i_2(r,p,\eta),$$
where
\begin{gather*}
    V^i_1(r,p,\eta)  =  \frac{1}{\sigma^3 \eta^{3/5}} \Big[ r g^i(p) + \frac{4\sqrt\lambda  m_{\ast}r^3}{3 f}\Big], \quad  V^i_2(r,p,\eta)  = -\frac{r p}{ 10 \eta}.
\end{gather*}
Since $V_1^i(r,p,\eta)=\mathcal O(\eta^{-3/5})\mathcal O(d^2)$ and $V_2^i(r,p,\eta)=\mathcal O(\eta^{-1})\mathcal O(d^2)$ as $\eta\to\infty$ and $d\to 0$, then for all $0<\varepsilon_1<1$ and $i\in\{1,2\}$ there exist $d_1>0$ and $\eta_1>0$ such that
\begin{gather}
\label{LFestAS}
(1-\varepsilon_1)\Big[\sqrt{\lambda} r^2+\frac{( m- m_{\ast}\cos\psi_0^i)f p^2}{2 m_{\ast}}\Big]\leq L^i(r,p,\eta)\leq (1+\varepsilon_1)\Big[\sqrt{\lambda} r^2+\frac{( m- m_{\ast}\cos\psi_0^i)f p^2}{2 m_{\ast}}\Big]
\end{gather}
for $(r,p,\eta)\in\mathcal D(d_1,\eta_1)$. The derivatives of $H^i(r,p,\eta)$, $V_1^i(r,p,\eta)$, and $V_2^i(r,p,\eta)$ with respect to $\eta$ along the trajectories of system \eqref{Sys0} have the following asymptotics as $\eta\to\infty$ and $d\to 0$:
\begin{eqnarray*}
    \frac{dH^i}{d\eta}\Big|_\eqref{Sys0} & = & \frac{g^i(p) }{\sigma^3 \eta^{3/5}} \Big[\frac{f  m}{2 m_{\ast}} \sin 2p -f \cos\psi_0^i \sin p\Big] - \\
     & & -\frac{ ( m- m_{\ast}\cos\psi_0^i) f p^2}{5 m_{\ast}\eta}+\mathcal O(\eta^{-1})\mathcal O(d^3) + \mathcal O(\eta^{-6/5})\mathcal O(d^2), \\
    \frac{dV^i_1}{d\eta}\Big|_\eqref{Sys0} &=&  - \frac{g^i(p) }{\sigma^3 \eta^{3/5}} \Big[\frac{f  m}{2 m_{\ast}}  \sin 2p -f \cos\psi_0^i \sin p\Big]+ \mathcal O(\eta^{-1})\mathcal O(d^3), \\
    \frac{dV^i_2}{d\eta}\Big|_\eqref{Sys0} &=& \frac{1}{5 \eta} \Big[ - \sqrt \lambda r^2 + \frac{ ( m- m_{\ast}\cos\psi_0^i)f p^2}{2 m_{\ast}} \Big] + \mathcal O(\eta^{-1})\mathcal O(d^3)+ \mathcal O(\eta^{-6/5})\mathcal O(d^2).
\end{eqnarray*}
These formulas imply the expression for the total derivative of function $L^i(r,p,\eta)$, which happens to have a sign-definite leading term in the asymptotics: \begin{eqnarray*}
    \frac{dL^i}{d\eta}\Big|_\eqref{Sys0} & = & -\frac{1}{5\eta} \Big[ \sqrt\lambda r^2 + \frac{ ( m- m_{\ast}\cos\psi_0^i) f p^2}{2 m_{\ast}} \Big] + \mathcal O(\eta^{-1})\mathcal O(d^3)+ \mathcal O(\eta^{-6/5})\mathcal O(d^2).
\end{eqnarray*}
Since the remainders in the latter expression can be made arbitrarily small, it follows that for all $0<\varepsilon_2<1$ there exist $d_2>0$ and $\eta_2>0$ such that
\begin{eqnarray}
\label{Lderiv}
    \frac{dL^i}{d\eta}\Big|_\eqref{Sys0} \leq -\frac{(1-\varepsilon_2)}{5 \eta} \Big[ \sqrt\lambda r^2 + \frac{ ( m- m_{\ast}\cos\psi_0^i) f p^2}{2 m_{\ast}} \Big]
\end{eqnarray}
for $(r,p,\eta)\in\mathcal D(d_2,\eta_2)$.
Thus for all $i\in\{1,2\}$ the constructed function $L^i(r,p,\eta)$ satisfies the inequalities
\begin{gather}
\label{LFest}
  (1-\varepsilon_1) A^i d^2 \leq L^i(r,p,\eta)\leq (1+\varepsilon_1) B^i d^2, \quad \frac{dL^i}{d\eta}\Big|_\eqref{Sys0}\leq - \frac{(1-\varepsilon_2)  A^i d^2}{5 \eta},\\
\nonumber   A^i:=\min\{\sqrt\lambda,( m- m_{\ast}\cos\psi_0^i)(2 m_{\ast})^{-1}f\}>0, \\
\nonumber   B^i:=\max\{\sqrt\lambda,( m- m_{\ast}\cos\psi_0^i)(2 m_{\ast})^{-1}f\}>0,
\end{gather}
for all $(r,p,\eta)\in\mathcal D(d_0,\eta_0)$, where $d_0:=\min\{d_1,d_2\}$ and $\eta_0:=\max\{\eta_1,\eta_2\}$.
Let $\epsilon$ be an arbitrary positive constant such that $0<\epsilon<d_0$, then
\begin{gather*}
   \sup_{d<\delta,\eta>\eta_0} L^i(r,p,\eta)\leq (1+\varepsilon_1)B^i \delta^2< (1-\varepsilon_1)A^i \epsilon^2 \leq  \inf_{d=\epsilon,\eta>\eta_0} L^i(r,p,\eta), \quad \delta:=\epsilon \Big[\frac{(1-\varepsilon_1)A^i}{2(1+\varepsilon_1)B^i}\Big]^{1/2}.
\end{gather*}
Hence, any solution $r(\eta)$, $p(\eta)$ of system \eqref{Sys0} with initial data $\sqrt{r^2(\eta_0)+p^2(\eta_0)} \leq \delta$
cannot leave $\epsilon$-neighborhood of the equilibrium $(0,0)$ as $\eta>\eta_0$. Therefore, for all $i\in\{1,2\}$ the equilibrium $(0,0)$ of system \eqref{Sys0} is stable as $\eta>\eta_0$. The stability on the finite interval $(0,\eta_0]$ follows from the theorem on the continuity of the solution to the Cauchy problem with respect to the initial data (see~\cite[Ch. 3]{Halil}).

Let us show that the equilibrium is asymptotically stable. Indeed, consider the solution $r(\eta)$, $p(\eta)$ to system \eqref{Sys0} with initial data $ r^2(\eta_0)+p^2(\eta_0) \leq d_0^2$, then for all $i\in\{1,2\}$ the function $\ell_i(\eta)=L^i\big(r(\eta),p(\eta),\eta\big)$ satisfies the following inequality as $\eta>\eta_0$:
$$\frac{d\ell_i}{d\eta}\leq - \frac{\beta \ell_i}{\eta}, \quad  \beta:=\frac{(1-\varepsilon_2)A^i}{5 (1+\varepsilon_1)B^i}, \quad 0<\beta<\frac{1}{5}.$$
Integrating the last expression with respect to $\eta$, we obtain:
$$0\leq \ell_i(\eta)\leq \ell_i(\eta_0) \exp\Big(-\beta\int\limits_{\eta_0}^\eta \frac{dz}{z }\Big) \leq B^i d_0^2 \cdot \Big(\frac{\eta}{\eta_0}\Big)^{-\beta} \quad {\rm as} \quad \eta>\eta_0.$$
Combining this with \eqref{LFestAS}, we get the asymptotic estimate $r^2(\eta) + p^2(\eta) = \mathcal O(\eta^{-\beta})$ as $\eta>\eta_0$.
By means of change of variables we derive the asymptotic estimates \eqref{AsEst} with $s=5\beta/8$, $\tau_0=\sigma^4\eta_0^{4/5}$ and $\delta_0=d_0$ for the solutions to system \eqref{PRst} with initial data from a neighborhood of the points $(\rho_\ast^i(\tau_0),\psi_\ast^i(\tau_0))$, $i\in\{1,2\}$. This completes the proof.
\endproof
\begin{Rem}
The constructed Lyapunov function $L^2(r,p,\eta)$ satisfies the inequalities \eqref{LFest} as $m>-m_{\ast}$. Therefore, the assertion of Theorem remains valid for the solution $\rho_\ast^2(\tau)$, $\psi_\ast^2(\tau)$ whenever  $m>-m_{\ast}$ and $m\neq m_\ast$.
\end{Rem}

We have the following theorem for another pair of solutions $\rho_\ast^3(\tau)$, $\psi_\ast^3(\tau)$ and $\rho_\ast^4(\tau)$, $\psi_\ast^4(\tau)$
\begin{Th}
\label{Th2}
If $m<-f/\sqrt{4\lambda}$, then the solutions $\rho_\ast^3(\tau)$, $\psi_\ast^3(\tau)$ and $\rho_\ast^4(\tau)$, $\psi_\ast^4(\tau)$ are asymptotically stable.  Moreover for all $i\in\{3,4\}$ and $0<s<1/8$ there exist $\delta_0>0$ and $\tau_0>0$ such that for all $(R_0, \Psi_0){\rm :}$ $ (R_0-\rho^i_\ast(\tau_0))^2+(\Psi_0-\psi^i_\ast(\tau_0))^2\leq \delta_0^2$ the solution $\rho(\tau)$, $\psi(\tau)$ to system \eqref{PRst} with initial data $\rho(\tau_0)=R_0$, $\psi(\tau_0)=\Psi_0$ has the asymptotics
\begin{gather}
\rho(\tau)=\sqrt{\lambda \tau}+\mathcal O(\tau^{-1/4-s}), \quad \psi(\tau)=\psi_0^i+\mathcal O(\tau^{-s}) \quad {\rm as }\quad \tau>\tau_0,
\label{AsEst34}
\end{gather}
where $\psi_0^3=\arccos( m_{\ast}/ m)$, $\psi_0^4=-\arccos( m_{\ast}/ m)$, and $m_{\ast}=f/\sqrt{4\lambda}$.
\end{Th}
\proof
In system \eqref{PRst} we make the change of variables $\rho(\tau)=\rho^i_\ast(\tau) + r(\eta) \tau^{-1/4}$, $\psi(\tau)=\psi_\ast^i(\tau)+p(\eta)$, $\eta=4\tau^{5/4}/5$ and we study the stability of the fixed point $(0,0)$ in the system \eqref{Sys0} with $i=3$ and $i=4$.
The construction of the Lyapunov function is based on the asymptotic behaviour of the right-hand sides of system \eqref{Sys0} as $\eta\to\infty$. We see that the functions $\partial_p H^i(r,p,\eta)$, $\partial_r H^i(r,p,\eta)$, and $G^i(r,p,\eta)$ have the following asymptotics
\begin{eqnarray*}
\partial_p H^i  &=& \frac{  (g^i(p))' f}{2 m_{\ast}} + \frac{f\psi_1^i}{\sigma^2 \eta^{2/5}} \Big[ (1-\cos p)\cos\psi_0^i +\frac{ m}{ m_{\ast}}  (\cos 2p - 1)\cos2\psi_0^i\Big] - \\
                & & -\frac{f\psi_1^i}{\sigma^2 \eta^{2/5}} \Big[\sin\psi_0^i \sin p+\frac{ m}{ m_{\ast}}  \sin 2\psi_0^i \sin 2p\Big]+ \mathcal O(\eta^{-3/5}) \mathcal O(d), \\
  \partial_r H^i &=& 2\sqrt\lambda r+\mathcal O(\eta^{-3/5}) \mathcal O(d), \\
                G^i &=&  \frac{g^i(p)}{\sigma^3 \eta^{3/5}}  - \frac{1}{5 \eta} [p + \mathcal O(d^2)]  + \mathcal O(\eta^{-6/5})\mathcal O(d),\\
           g^i (p)&:=& \frac{ m}{2} [\cos 2\psi_0^i-\cos(2p+2\psi_0^i)]-2 m_{\ast}[\cos\psi_0^i-\cos(p+\psi_0^i)]=\mathcal O(d^2),
\end{eqnarray*}
as $\eta\to \infty$ and $d\to 0$. In the case $m<-f/\sqrt{4\lambda}$, we can select in the Hamiltonian $H^i(r,p,\eta)$ a positive definite quadratic form as the main term of the asymptotics:
\begin{eqnarray*}
 H^i   &=& \sqrt\lambda r^2  + \frac{fm_0}{4 m_{\ast}}\Big[(1-\cos 2p)\cos2\psi_0^i+(\sin2p-2p)\sin2\psi_0^i \Big]- \\
                 & & - f\Big[ (1-\cos p)\cos\psi_0^i + (\sin p-p)\sin\psi_0^i\Big]+\\
                & & +\frac{f \psi_1^i}{\sigma^2 \eta^{2/5}}\Big[ (p- \sin p)\cos\psi_0^i  +  (\cos p -1)\sin\psi_0^i  \Big]+ \\
                & & +\frac{f \psi_1^i  m}{2 m_{\ast}\sigma^2 \eta^{2/5}}\Big[  (\sin 2p-2p)\cos 2\psi_0^i + (\cos2p-1)\sin2\psi_0^i  \Big]+ \mathcal O(\eta^{-3/5})\mathcal O(d^2)=\\
                & =& \sqrt\lambda r^2 + \frac{  ( m_{\ast}^2- m^2) f p^2}{2 mm_{\ast}}+\mathcal O(d^3)+\mathcal O(\eta^{-2/5})\mathcal O(d^2),
\end{eqnarray*}
as $\eta\to\infty$ and $d\to 0$. The basis of the construction of the Lyapunov function is the Hamiltonian perturbed by additional terms decreasing as $\eta\to\infty$ with different rates: $L^i(r,p,\eta) = H^i(r,p,\eta) + V^i_1(r,p,\eta)+ V^i_2(r,p,\eta)$, where
\begin{gather*}
    V^i_1(r,p,\eta)  =  \frac{1}{\sigma^3 \eta^{3/5}} \Big[ r g^i(p) + \frac{4\sqrt\lambda  m_{\ast}r^3}{3 f}\Big], \quad  V^i_2(r,p,\eta)  = -\frac{r p}{ 10 \eta}.
\end{gather*}
It is easy to see that for all $0<\varepsilon_1<1$ and $i\in\{3,4\}$ there exist $d_1>0$ and $\eta_1>0$ such that
\begin{gather*}
(1-\varepsilon_1)\Big[\sqrt{\lambda} r^2+\frac{ ( m_{\ast}^2- m^2) f p^2}{2 mm_{\ast}}\Big]\leq L^i(r,p,\eta)\leq (1+\varepsilon_1)\Big[\sqrt{\lambda} r^2+\frac{ ( m_{\ast}^2- m^2) f p^2}{2 mm_{\ast}}\Big]
\end{gather*}
for $(r,p,\eta)\in\mathcal D(d_1,\eta_1)$. The total derivative of the Hamiltonian $H^i(r,p,\eta)$ and the additional terms $V^i_k(r,p,\eta)$ have the following asymptotics:
\begin{eqnarray*}
     \frac{dH^i}{d\eta}\Big|_\eqref{Sys0}   & = &  \frac{ (g^i(p))' g^i(p) f}{2 m_{\ast}\sigma^3 \eta^{3/5}} - \frac{ ( m_{\ast}^2- m^2) f p^2}{5  m m_{\ast}\eta}+\mathcal O(\eta^{-1})\mathcal O(d^3) + \mathcal O(\eta^{-6/5})\mathcal O(d^2),\\
    \frac{dV_1^i}{d\eta}\Big|_\eqref{Sys0}  & = &  - \frac{ (g^i(p))'g^i(p)f }{2 m_{\ast}\sigma^3 \eta^{3/5}} + \mathcal O(\eta^{-1})\mathcal O(d^3), \\
    \frac{dV^i_2}{d\eta}\Big|_\eqref{Sys0} &=& \frac{1}{5 \eta} \Big[ - \sqrt \lambda r^2 + \frac{ ( m_{\ast}^2- m^2) f p^2}{2 mm_{\ast}} \Big] + \mathcal O(\eta^{-1})\mathcal O(d^3)+ \mathcal O(\eta^{-6/5})\mathcal O(d^2).
\end{eqnarray*}
This implies that the derivative of the function $L^i(r,p,\eta)$ with respect to $\eta$ along the trajectories of system \eqref{Sys0} consists a quadratic form in the leading term of its asymptotic expansion:
\begin{eqnarray*}
    \frac{dL^i}{d\eta}\Big|_\eqref{Sys0} & = & - \frac{1}{5 \eta} \Big[ \sqrt \lambda r^2 + \frac{ ( m_{\ast}^2- m^2) f p^2}{2 mm_{\ast}} \Big] + \mathcal O(\eta^{-1})\mathcal O(d^3)+ \mathcal O(\eta^{-4/3})\mathcal O(d^2).
\end{eqnarray*}
Hence, for all $0<\varepsilon_2<1$ and $i\in\{3,4\}$ there exist $d_2>0$ and $\eta_2>0$ such that
\begin{eqnarray*}
    \frac{dL^i}{d\eta}\Big|_\eqref{Sys0} \leq -  \frac{(1-\varepsilon_2)}{5 \eta} \Big[ \sqrt \lambda r^2 + \frac{ ( m_{\ast}^2- m^2) f p^2}{2 m m_{\ast}} \Big]
\end{eqnarray*}
for $(r,p,\eta)\in\mathcal D(d_2,\eta_2)$. Thus, for all $i\in\{3,4\}$ the Lyapunov function $L^i(r,p,\eta)$ satisfies the inequalities  \eqref{LFest} with $A^i:=\min\{\sqrt\lambda, ( m_{\ast}^2- m^2)f p^2/2 mm_{\ast}\}$ and $B^i:=\max\{\sqrt\lambda, ( m_{\ast}^2- m^2)f p^2/2 mm_{\ast}\}$ in the domain $\mathcal D(d_0,\eta_0)$, where $d_0:=\min\{d_1,d_2\}$ and $\eta_0:=\max\{\eta_1,\eta_2\}$. This implies that the equilibrium $(0,0)$ of system \eqref{Sys0} is stable. By integrating the second estimate in \eqref{LFest} we obtain the asymptotic stability of the equilibrium. The change-of-variables formula implies the stability of the solutions $\rho_\ast^i(\tau)$, $\psi_\ast^i(\tau)$, $i\in\{3,4\}$ and the asymptotic estimates \eqref{AsEst34} with $\tau_0=\sigma^4\eta_0^{4/5}$ and $\delta_0=d_0$ for the solutions to system \eqref{PRst} with initial data from the $\delta_0$-neighbourhood of the points $(\rho_\ast^i(\tau_0),\psi_\ast^i(\tau_0))$, $i\in\{3,4\}$.
\endproof

The table~\ref{Tab2} shows the final results of our stability analysis. We see that unstable autoresonant solution $\rho_\ast^1(\tau)\sim\sqrt{\lambda\tau}$, $\psi_\ast^1(\tau)\sim 0$ to system \eqref{PRst} is stabilized with increasing of values $m$, while the stability of the solution $\rho_\ast^2(\tau)\sim\sqrt{\lambda\tau}$, $\psi_\ast^2(\tau)\sim \pi$ is not destroyed.
On the other hand, the decreasing of the parameter $m$ leads to stabilization of another pair of captured solutions $\rho_\ast^3(\tau)$, $\psi_\ast^3(\tau)$ and $\rho_\ast^4(\tau)$, $\psi_\ast^4(\tau)$. Note that the property of asymptotic stability implies that each stable autoresonant solution is the attractor for two-parametric family of captured solutions to system \eqref{PRst}.
At the same time, from asymptotic estimates \eqref{AsEst} and \eqref{AsEst34} it follows that we have only power convergence to the isolated solutions with asymptotics \eqref{ASstep}. Proving the absence of the exponential convergence involves the asymptotic analysis of two-parametric family of general solutions to system \eqref{PRst}. This will be discussed in another paper.

\begin{table}[H]
\caption{Stability of solutions $\rho_\ast^i(\tau)$, $\psi_\ast^i(\tau)$}
\label{Tab2}
\begin{tabular}{c|c|c|c}
\hline         & $ m<- m_{\ast}$ & $- m_{\ast}< m< m_{\ast}$ & $ m> m_{\ast}$  \\
\hline
$i=1$ & unstable  & unstable & stable \\
\hline
$i=2$ & unstable  & stable & stable \\
\hline
$i=3$ & stable  & -  & unstable \\
\hline
$i=4$ & stable  &  -  & unstable \\
\hline
\end{tabular}
\end{table}

\section{Conclusion}

The proposed theory for the model system \eqref{PRst} can be used in the study of autoresonance phenomena in nonlinear oscillatory systems described by equation \eqref{EPst}. Let $\mu=\varepsilon\alpha^{-3/4}$ and $\mu_0=(16h_0^2/3\gamma f_0^2)^{3/4}$. From Theorems \ref{Th1} and \ref{Th2} it follows that if $\mu>\mu_0$, then there exist a stable autoresonant solution with the asymptotics
\begin{gather}
\label{as}
u(t)=\varepsilon^{1/3} \rho(\varepsilon^{2/3} t ) \sqrt{\frac{8}{3\gamma}}  \cos \Big(\psi(\varepsilon^{2/3} t)-\phi(t)\Big)+\mathcal O(\varepsilon), \quad 0\leq t\leq \mathcal O(\varepsilon^{-2/3}),
\end{gather}
where $\rho(\tau)=\sqrt{\lambda\tau}+o(\tau^{-1/4})$, $\lambda=2\alpha\varepsilon^{-4/3}$ and $\psi(\tau)=\pi+o(1)$ as $\tau\to\infty$.
If $0<\mu<\mu_0$,  then there exist a pair of stable autoresonant solution with the asymptotics \eqref{as}, where $\rho(\tau)=\sqrt{\lambda\tau}+o(\tau^{-1/4})$, $\psi(\tau)=\psi_0+o(1)$ as $\tau\to\infty$. In this case, $\psi_0\in\{0,\pi\}$ if $h_0>0$, and $\psi_0\in\{\pm \arccos (\mu/\mu_0)^{4/3}\}$ if $h_0<0$.

In summary, we have investigated the stabilization problem of unstable autoresonant modes in nonlinear oscillatory systems with the external pumping by small adiabatically varying parametric perturbation. We have considered the isolated  autoresonant solutions with power asymptotics at infinity and have investigated the dependence of their stability on the values of the perturbation parameters. It was shown that the unstable autoresonant solutions become stable as the parameter $m/m_\ast=m\sqrt{4\lambda f^{-2}}$ passes through the critical values $-1$ and $1$. The behaviour of captured solutions at the bifurcation points has not been considered here. This will be discussed elsewhere.

\end{document}